# Empowering Learning: Standalone, Browser-Only Courses for Seamless Education

**Babak Moghadas**[1]
bmoghad1@jhu.edu

**Brian S. Caffo**[1,2]
bcaffo1@jhu.edu

[1] Bloomberg School of Public Health, Johns Hopkins University

[2] Department of Biomedical Engineering, Johns Hopkins University

## Abstract:

Massive Open Online Courses (MOOCs) have transformed the educational landscape, offering scalable and flexible learning opportunities, particularly in data-centric fields like data science and artificial intelligence. Incorporating AI and data science into MOOCs is a potential means of enhancing the learning experience through adaptive learning approaches. In this context, we introduce PyGlide, a proof-of-concept open-source MOOC delivery system that underscores autonomy, transparency, and collaboration in maintaining course content. We provide a user-friendly, step-by-step guide for PyGlide, emphasizing its distinct advantage of not requiring any local software installation for students. Highlighting its potential to enhance accessibility, inclusivity, and the manageability of course materials, we showcase PyGlide's practical application in a continuous integration pipeline on GitHub. We believe that PyGlide charts a promising course for the future of open-source MOOCs, effectively addressing crucial challenges in online education.

# 1. Introduction:

Massive Open Online Courses (MOOCs) have emerged as a revolutionary form of online education, transforming the traditional learning landscape. MOOCs are typically web-based platforms that offer open-access or low-cost access of courses and programs to a global audience, thereby providing unprecedented access to high-quality educational content from experts (Deng et al., 2019). While differing in enrollment numbers and openness, the primary hallmark of MOOCs is their scalability, enabling thousands, if not millions, of students to enroll in a course or program simultaneously. However, from a learner's perspective, a key advantage of MOOCs is their flexibility, as learners can access the course materials at their own pace and convenience. This asynchronous learning model caters to a diverse range of learners, from working professionals seeking to upskill or change careers, to students looking to supplement their formal education. Moreover, MOOCs often offer a wide array of subjects, spanning disciplines such as computer science, business, humanities, and beyond, allowing learners to explore their interests beyond traditional educational boundaries.

Data oriented topics, such as: data-oriented programming, data science, data analysis, statistics, machine learning and artificial intelligence (AI), have been among the most successful and subscribed content in MOOC instruction. These offer a unique combination of being highly sought after for upskilling, working well in online and asynchronous delivery, and being growth careers at the same time MOOCs were developing as an instructional medium.

A recent crucial aspect of MOOCs is their interactive nature and focus on active learning. Many platforms incorporate multimedia elements, including video lectures, quizzes, and coding assignments, all fostering active engagement and knowledge retention. Additionally, discussion forums and social learning features encourage students to collaborate, exchange ideas, and seek help from peers, creating a sense of virtual community and support. Importantly, most MOOCs are delivered in a browser, thus with few system requirements for learners. Many platforms focusing on data science include web access to computational and development environments that abstract installation of software and libraries. This is ideal for novice learners allowing for focus on core concepts over IT issues. In addition, it offers a core benefit to instructors and learners of a highly consistent pedagogical environment.

The integration of AI and data science into MOOC instruction and delivery could further enhance the learning experience. AI and analytically informed adaptive learning approaches could ensure that learners receive tailored support and challenges, maximizing their understanding and mastery of the subject matter. The continuous evolution of MOOCs, and the integration of AI and data science hold tremendous potential for the future of instruction in MOOC appropriate areas of education.

A benefit, and yet also possible concern, with MOOCs is the necessary connection with one or more centralized delivery entities. This includes single instructors, universities and MOOC content providers, such as Coursera, EdX, Udacity and Udemy. While these collaborations have been highly successful, open-source MOOCs, following the open-source software model, have a place to fill in content delivery in the same way open-source software and community knowledge bases, like Wikipedia, have a role along with their more centralized counterparts. The goal of this manuscript is to present a proof-of-concept open-source MOOC delivery system.

Open-source MOOCs represent an evolution of scalable online education. These MOOCs incentivize collaboration, innovation and shared ownership within the educational community.

Unlike traditional MOOC platforms, open-source MOOCs prioritize autonomy, transparency, and collaboration.

In an open-source MOOC, course content is maintained by the community and is deployed in an open-source fashion, ideally on open-source software. In a traditional MOOC, the instructor and the platform dictate course offerings and policies, which offers consistency and quality control, but can also limit the diversity of educational opportunities. In an open-source MOOC, the boundaries between instructors, learners and content are blurred so that all can modify, and reuse content as needed without oversight. In addition, learners have greater control over their own data, learning progress and privacy.

Ideally, open-source MOOCs would be serverless yet still offer the same continuity of experience as centralized cloud-based MOOC platforms, since the cost of compute environments generally will need to be centralized. Of course, some method for distributing software is needed so that a minimal, yet convenient publication would only require web hosting, but would not require video hosting services or computational servers for interactive exercises. Moreover, web hosting should be performed so that learners and instructors can fork and modify content with ease. The cost of deploying and engaging in an open-source MOOC should be minimal. In what follows, we demonstrate a proof-of-concept MOOC delivery system which we call PyGlide. This combines web assembly-based exercises along with audio and slide content as well as large language model integration.

## 2. Design Philosophy

Creating online courses is a rewarding endeavor, but it also comes with its own set of challenges and issues, especially course maintenance and updating material (Candace Savonen Carrie Wright & Leek, 2023; Wilson, 2016). Ensuring the course remains accurate and valuable over time is the main challenge for any MOOC instructor or the creator in a rapidly evolving field like data science. Even if core concepts remain constant core applications may significantly vary over time. To alleviate these issues, we propose a python library to generate interactive learning slides as HTML that only call client-side scripts. Delivered in this manner, the library will require no setup from students other than clicking on a link and will also require minimum efforts from instructor to publish and maintain the course. We refer to the library as Python Guided Learning through Interactive Digital Education (PyGlide). PyGlide will improve accessibility and inclusivity of the course materials by reducing entry barriers. The inclusivity fosters a more equitable learning environment, empowering individuals who may have faced obstacles in traditional educational settings. In addition, it will ease out the maintainability of the course for instructor.

## 3. Structure and elements:

Below we outline the main structural elements of our library PyGlide. Briefly, we use web assembly based [PyScript](#) and [Jupyter notebooks](#) combined with text to speech (TTS) audio overlays compiled with [Nbconvert](#) to create online pseudo-courses.

### 3.1. Jupyter Notebook

The advent of Jupyter Notebooks has significantly transformed the landscape of data analysis and visualization. Jupyter Notebooks have gained immense popularity in recent years, becoming a standard in data-driven research and analysis. Unlike traditional code editors, Jupyter Notebooks combine executable code with multimedia elements, such as text, images, and audio. This unique approach makes it a suitable tool for creating educational content in a self-contained and maintainable manner. In addition, Jupyter Notebooks offer a dynamic platform for data exploration and visualization, enabling researchers to interactively inspect datasets and generate visual representations. There are some additional tools like Nbconvert that can convert and produce file formats that are more common for data presentation. Nbconvert is a versatile tool within the Jupyter ecosystem that can convert Jupyter Notebooks into various output formats, including HTML, PDF, and LaTeX. Leveraging nbconvert, one can create content that embed executable code blocks, interactive plots, and live data.

### 3.2. PyScript

Active learning is a paradigm where one actively engages in material during content delivery. Frequent exercises amid lecture content are known to be preferable to larger chunks of lecture content followed by exercises (Lang, 2021). In data science this translates to actively performing data science tasks during instruction. However, switching between content and an integrated development environment (IDE) introduces friction. For example, it requires installing and maintaining a data-science scripting language, we focus on python, or a centralized platform hosting an IDE as a service. Python installation, in particular, can be challenging for learners with limited computational backgrounds. Hosted solutions are expensive or require reliance on third party offerings, which may not continue to offer free or low-cost solutions in the future. PyScript offers a seamless solution to execute Python code directly within the web browser. With PyScript, one can easily integrate an interactive Python REPL into a website, distribute interactive dashboards by sharing them as HTML files, and even develop dynamic client-side web applications powered by Python. PyScript execution inside HTML input allows students to experiment with the presented code, altering parameters, and observing real-time changes.

### 3.3. Text-to-Speech (TTS)

Companion audio content along with slide or lecture content is desirable for many reasons. Foremost, audio content allows learners with visual impairments or reading challenges to access material in a preferable form. In fact, the increase in access to educational materials is often cited as one of the main benefits of MOOCs (Kopp et al., 2017). However, audio and video content are the most difficult aspect of a MOOC to maintain. Integrating Text-to-Speech (TTS) technology can bridge this gap by providing an audio representation of the course material, thereby making the learning experience more accessible and inclusive. In addition, combining visual and auditory modalities through TTS encourages multisensory learning, leading to improved retention and understanding of course content. Rapid advances in open-source TTS libraries make their use more compelling. Importantly, TTS solutions bring no further copyright restrictions beyond those imposed by the web hosting.

### 3.4. AI support:

The one-size-fits-all approach of traditional MOOCs does not cater to all individual learning needs and preferences of diverse learners. AI tutors can play a significant role in the learning process. Large Language Models (LLM) represent a potentially accessible tutoring experience. Utilizing chatbots can provide a promising approach to enhance the effectiveness of MOOCs by offering continuous support throughout a learner's journey. These mentors are designed to engage learners in interactive conversations, answer questions, and recommend relevant resources. Of note, the most popular LLMs are particularly useful in data-science and python in specific, having been trained on a large collection of relevant text content.

## 4. Preparation and Usage

The `PyGlide` package is available on Pypi.org and can be installed from the command line using `pip install pyglide`.

```
> pyglide
usage: pyglide [-h] [-v] [-m] [-p] [-i INPUT] [filename]

positional arguments:
  filename          Pass the file name without extension

options:
  -h, --help        show this help message and exit
  -v, --version     Display the version of PyGlide
  -m, --mute        Mute the audio
  -p, --dPrompt     Disable prompt window
  -i INPUT          The file name without extension
```

Figure 1: Command line interface of PyGlide.

Currently, PyGlide provides some configuration options to customize the output slides by arguments when creating an instance of the `pyglide` class or running pyglide from the command line. Figure 1. shows the command line interface of the PyGlide. The available configuration options:

- `audio`: Specify if you wish the output without audio (default: "un-mute"). If you wish to generate the output without audio, you can pass "-m" as the value for this argument.
- `AI_assistant`: Specify if the output file should include an AI assistant (default: "AI prompt is included"). If you wish to generate the output without AI assistant, you can pass "-p" as the value for this argument.

The package is deployed with an example template ("original_example.ipynb") to help the user set up their content. The user can access the example notebook after installing PyGlide through command line by `pyglide original_example`. This will generate a folder `output` which will contain the example notebook file as shown in Figure 2.

```
> tree
.
├── original_example.ipynb
├── original_example_pyglide.html
└── slides_audios
    ├── DevelopingDataProducts.mp3
    └── ShinyApplication.mp3

2 directories, 4 files
```

Figure 2: The output of example template in current working directory.

The following step by step guideline, which is also available in package documentation file, will assist the user to create their interactive slides using pyglide.

1. Open a Jupyter notebook in Jupyter Lab or Jupyter Notebook.
2. Add a markdown cell at the top of the notebook.
3. Use level 1 heading to specify the title of the presentation and level 2 heading to specify the author and additional information (Figure 3).

```
1  # Developing Data Products
2  ### Presented by Brian Caffo
3  ### Date: 06/15/2023
```
Slide Type: slide   markdown

Figure 3: Setting the title of the presentation/ course, presenter's name and date.

4. Make sure each cell is (json) labeled with the correct Slide Type (e.g., Slide, Sub-Slide, Fragment, Skip, Notes, etc.). Slide Type notes is responsible for audio generation. This step is very important as it will determine the slides and the audio files that will be generated (Figure 4).

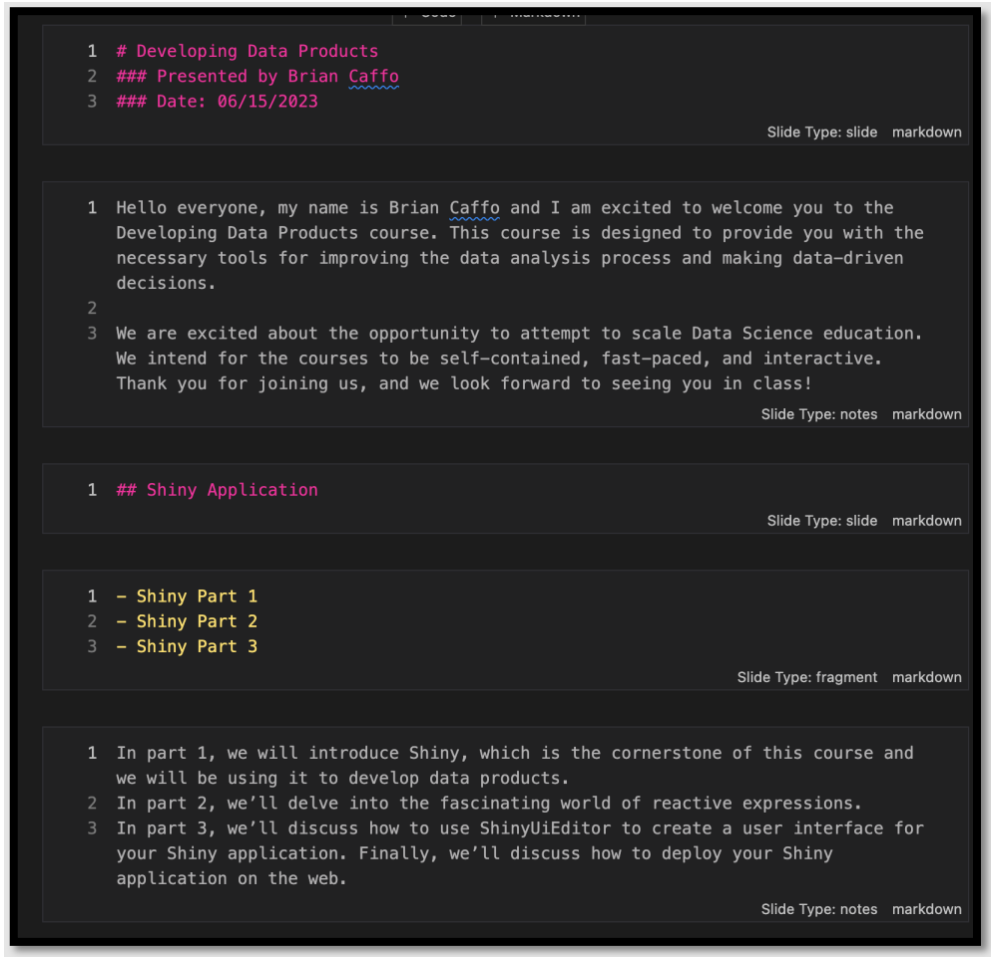

Figure 4: Accurately labeled cells with Slide Type to generate audio and slides

5. While preparing the slides you may wish to receive input from the user. PyGlide allows you to receive input from the user as simple text or as executable code. To do so, you should use `<div><!--Course_Text--></div>` in any slides you want to receive simple text. This HTML line will not be visible in the output slides. It will serve as a target. Similarly, to receive executable code, you can use `<div><!--Course_Code--></div>` in any slides you want to receive executable code. Figure 5 shows an example of implementing these labels. The package will automatically convert the input cells into interactive cells in the output slides.

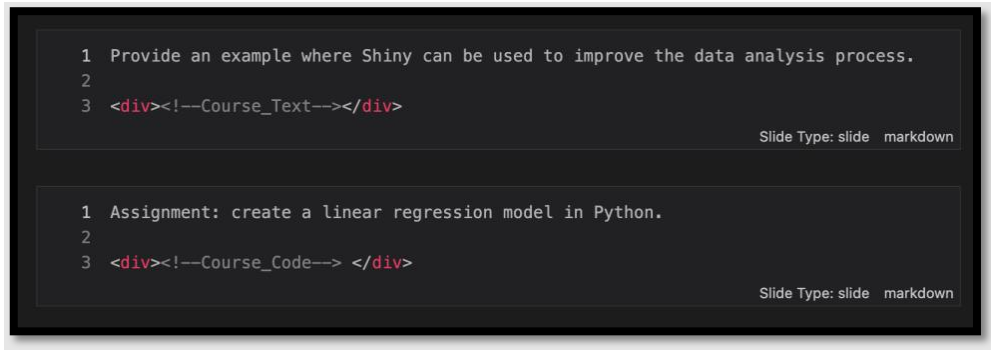

Figure 5: Labeling the cells to receive text input or executable code.

In the following few pictures, we are displaying the results of the original example template after compilation (Figure 6-10). The python code used in this example is provided below.

```python
import numpy as np
from sklearn.linear_model import LinearRegression
import matplotlib.pyplot as plt

# Generate more sample data
X = np.array([1, 2, 3, 4, 5, 6, 7, 8, 9, 10]).reshape(-1, 1)  # Input feature
y = np.array([2, 4, 5, 4, 5, 8, 9, 10, 11, 13])  # Target variable

# Create a linear regression model
model = LinearRegression()

# Fit the model to the data
model.fit(X, y)

# Create a denser range of test values
X_test = np.linspace(1, 10, 1000).reshape(-1, 1)

# Make predictions
y_pred = model.predict(X_test)

# Plot the data and regression line with more data points
plt.scatter(X, y, color='blue', label='Data Points')
plt.plot(X_test, y_pred, color='red', label='Regression Line')
plt.xlabel('Input Feature')
plt.ylabel('Target Variable')
plt.legend()
display(plt)
```

Listing 1: Python code example

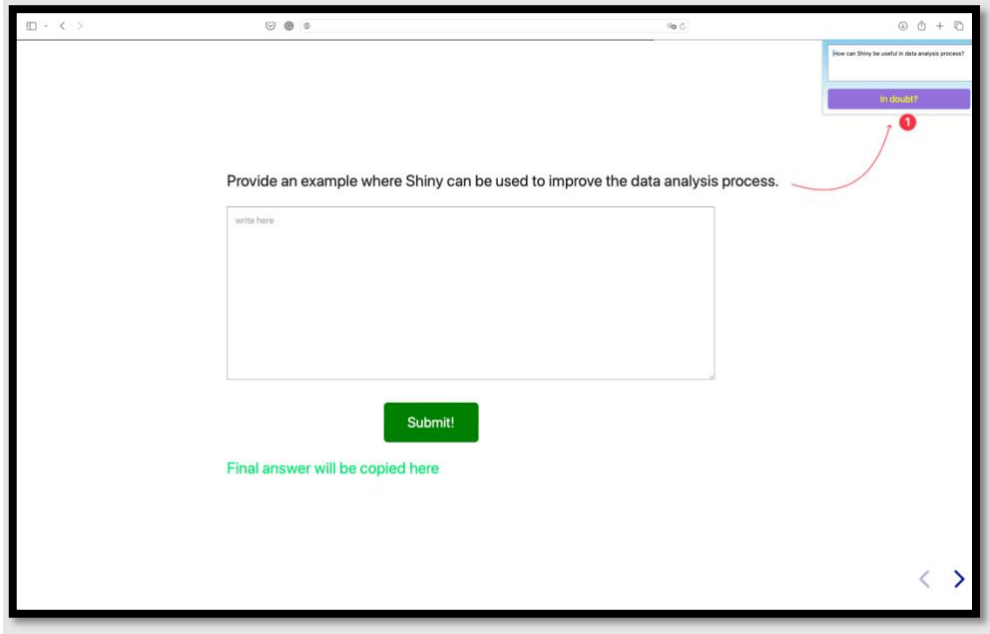

Figure 6: The interface for text input shown with the AI assistant at the top right corner.

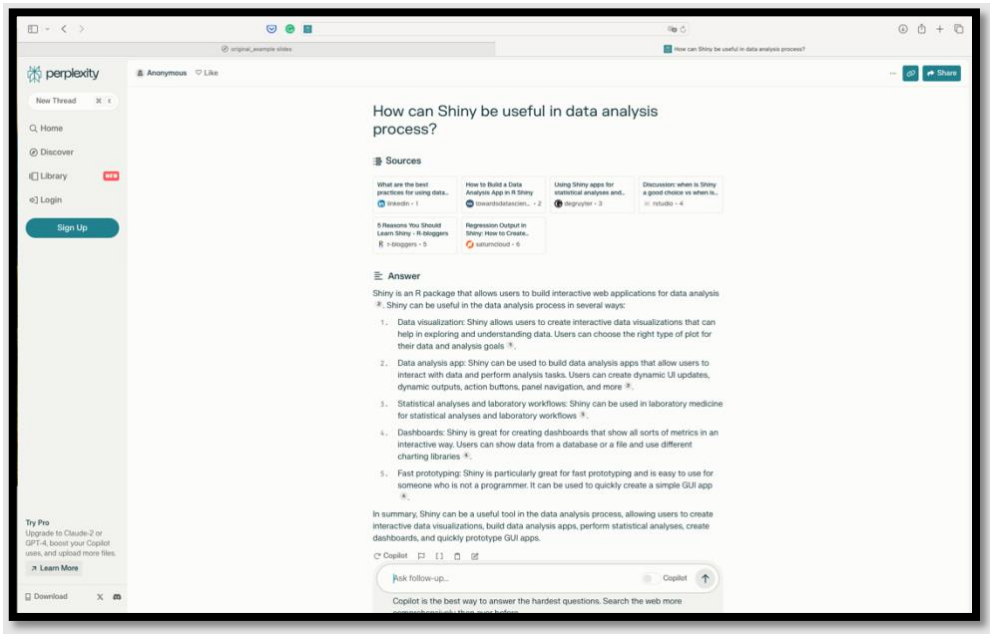

Figure 7: The result of the AI assistant for student's question.

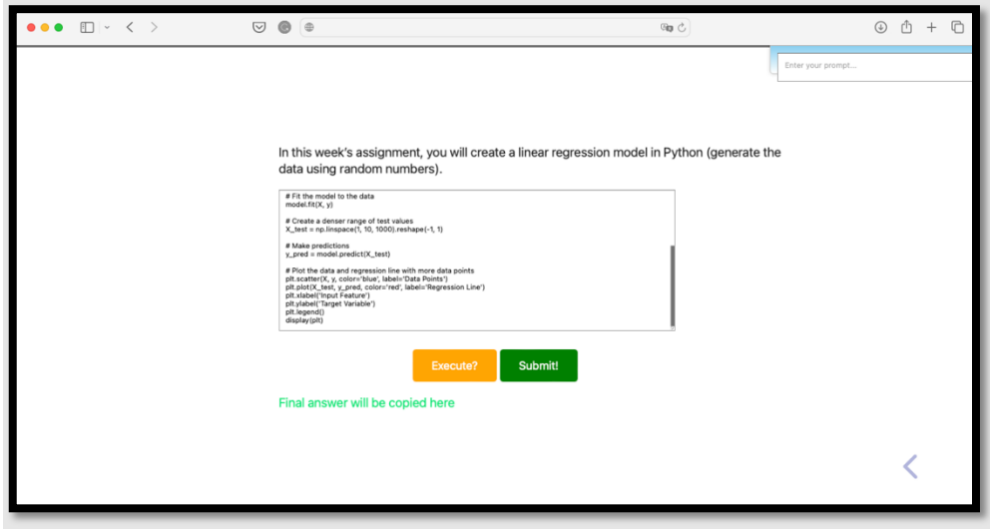

Figure 8: Interface for writing python code to practice or completing homework/ assignment.

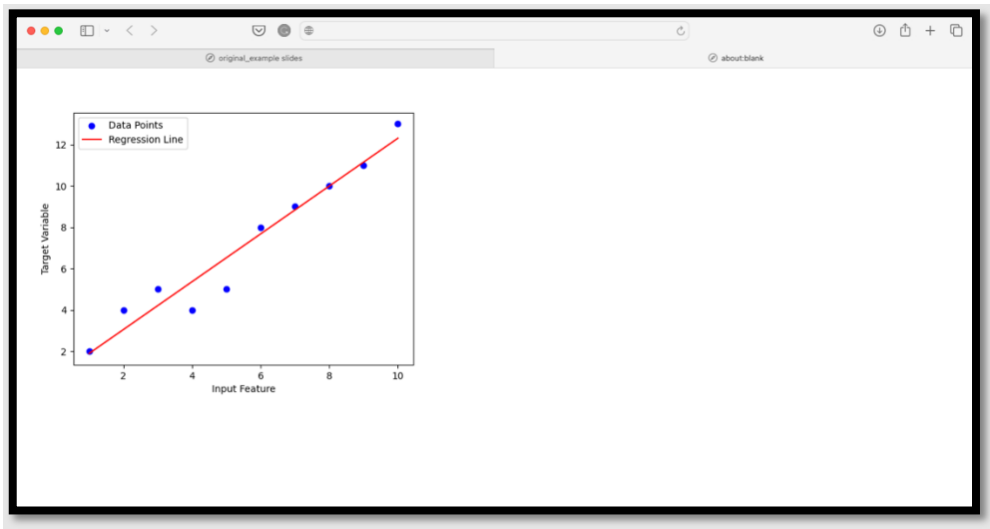

Figure 9: The executed result of the python code in PyScript.

We also created a real example course using continuous integration pipeline with PyGlide_Actions, which is a repository template on GitHub. Where it generates and deploys the course content automatically on GitHub pages. To use it, you can navigate to the template page and click "Use this template" and create your own repository, then enable GitHub pages setting, finally upload your notebooks to the repository and will run PyGlide, converts the notebooks to HTML and deploys them to GitHub pages providing you with the links to share with your audience.

## 5. Discussion:

Open-source MOOCs hold great promise in transforming the landscape of online education by promoting autonomy, data privacy, collaboration, and inclusivity. By leveraging decentralized technologies, these platforms offer a potential solution to the challenges faced by centralized MOOCs, empowering learners and educators with greater control over their educational journey. As AI mentorship continues to evolve, it has the potential to revolutionize the future of MOOCs, making quality education accessible and effective for learners worldwide. As a proof-of –principle, we introduced the PyGlide library to create standalone educational content powered by TTS and AI mentors. The PyGlide library offers a user-friendly method for transforming Jupyter notebooks into interactive slides with audio overlays. The use of TTS can reduce the efforts to maintain and update the content, since all the updates will automatically be implemented in the final HTML including the audio upon updating the text.

The PyGlide package leaves several components to be more fully developed. Ideally, the text input to the Pyscript prompts would include python IDE elements, such as autocompletion of methods and variables. In addition, stylistic changes would help engage learners further. However, the most glaring deficit is the ability to give credentials. Creating a centralized database of credential earners somewhat defeats the purpose of an open-source MOOC. One idea is to have credentials stored on a decentralized database, such as a blockchain, peer to peer sharing mechanism, such as a torrent or distributed filesystem such as IPFS. Interestingly, if a solution is used for decentralized storage of credentials, it could be used for content hosting as well, eliminating the need for centralized web hosting services or instructor deployed web servers. Finally, future developments would ideally include video overlays as well as audio overlays. This presents a problem as video hosting becomes storage intensive making it less viable for low resource hosting and forking or modification by others.